%% file: article.tex
\documentclass[11pt]{article}
\usepackage[margin=1in]{geometry}
\usepackage[utf8]{inputenc}
\usepackage[T1]{fontenc}
\usepackage{newtxtext,newtxmath}
\usepackage{microtype}
\usepackage{setspace}
\usepackage{graphicx}
\usepackage{booktabs}
\usepackage{array}
\usepackage{tabularx}
\usepackage{longtable}
\usepackage{adjustbox}
\usepackage{amsmath}
\usepackage{tikz}
\usetikzlibrary{arrows.meta,positioning,shapes.geometric,calc}
\usepackage[most]{tcolorbox}
\usepackage{caption}
\usepackage{subcaption}
\usepackage{float}
\usepackage{placeins}
\usepackage{xcolor}
\usepackage{hyperref}
\usepackage{enumitem}
\usepackage{pdflscape}
\hypersetup{colorlinks=true,linkcolor=blue!45!black,citecolor=blue!45!black,urlcolor=blue!45!black}
\definecolor{boxblue}{HTML}{EDF3FA}
\definecolor{boxgreen}{HTML}{EEF7EF}
\definecolor{boxorange}{HTML}{FFF5EA}
\definecolor{boxred}{HTML}{FDEDEC}
\newtcolorbox{findingbox}[1]{colback=boxblue,colframe=blue!45!black,title=#1,arc=2mm,boxrule=.5pt,left=2mm,right=2mm,top=1mm,bottom=1mm}
\newtcolorbox{methodbox}[1]{colback=boxgreen,colframe=green!45!black,title=#1,arc=2mm,boxrule=.5pt,left=2mm,right=2mm,top=1mm,bottom=1mm}
\newtcolorbox{cautionbox}[1]{colback=boxorange,colframe=orange!70!black,title=#1,arc=2mm,boxrule=.5pt,left=2mm,right=2mm,top=1mm,bottom=1mm}
\newtcolorbox{resultbox}[1]{colback=boxred,colframe=red!55!black,title=#1,arc=2mm,boxrule=.5pt,left=2mm,right=2mm,top=1mm,bottom=1mm}
\newcommand{\pp}{\mathrm{pp}}
\title{Stabilization without Inclusive Development:\\Neoliberalism, Economic Liberalization, Poverty, and Inequality in Bolivia}
\author{Ricardo Alonzo Fernandez Salguero\\\href{https://orcid.org/0000-0002-4189-961X}{ORCID: 0000-0002-4189-961X}}
\date{}
\begin{document}
\maketitle
\begin{abstract}
This article reconstructs the economic and social history of Bolivian neoliberalism and evaluates whether economic liberalization reduced or increased poverty and inequality in Bolivia. The historical argument is that the Bolivian neoliberal cycle was not a single event but a layered sequence: hyperinflation and emergency stabilization, the 1985 New Economic Policy, labor displacement and mining restructuring, second-generation reform in the 1990s, capitalization, decentralized state restructuring, commodity dependence, and the social conflicts that culminated in the collapse of the party system. The empirical contribution is to integrate macroeconomic indicators, economic-freedom indices, poverty and inequality series, IMF and financial-reform data, commodity and disaster controls, Bolivian export aggregates, and harmonized historical survey indicators. The preferred design is a heterogeneous instrumental-variables model that instruments domestic liberalization with lagged regional leave-one-out policy diffusion and allows Bolivia to differ from the Latin American average. The central estimate is that a 10-point increase in the Heritage economic-freedom score is associated, for Bolivia, with approximately +4.46 percentage points of poverty at the USD 4.20/day line, +3.61 percentage points at the USD 3/day line, +7.40 percentage points at the USD 8.30/day line, and +3.91 Gini points. These results remain socially regressive in sign after adding export-structure controls to the poverty specifications, although the causal interpretation remains conditional on the exclusion restriction. The article therefore advances a qualified conclusion: Bolivian neoliberalism stabilized hyperinflation, but the historically specific liberalization package appears to have increased social vulnerability and inequality rather than producing inclusive development.
\end{abstract}

\noindent\textbf{DOI:} \href{https://doi.org/10.5281/zenodo.20969752}{10.5281/zenodo.20969752}\\
\textbf{ORCID:} \href{https://orcid.org/0000-0002-4189-961X}{0000-0002-4189-961X}

\begin{cautionbox}{How to cite}
Fernandez Salguero, Ricardo Alonzo. (2026). \textit{Stabilization without Inclusive Development: Neoliberalism, Economic Liberalization, Poverty, and Inequality in Bolivia}. Zenodo. DOI: \href{https://doi.org/10.5281/zenodo.20969752}{10.5281/zenodo.20969752}.
\end{cautionbox}

\section{Introduction}

Bolivia is one of the most revealing cases for studying neoliberalism because the country experienced both an undeniable macroeconomic emergency and a socially conflictive restructuring process. By 1985 money, wages, prices, fiscal balances, and expectations had become unstable in a historically extreme way. The New Economic Policy launched through Supreme Decree 21060 reorganized exchange-rate policy, public finances, price regulation, labor relations, and state-enterprise governance \cite{ds21060,sachs_morales,cepal_bolivia}. The stabilization of hyperinflation was a real achievement. Yet stabilization is not identical to inclusive development. The same policy sequence that restored monetary order also displaced workers, weakened organized labor, restructured mining, opened sectors to private and foreign capital, and generated a new conflict over sovereignty, rents, services, and territory \cite{kohl2002,kohl_farthing2006,perreault2006,spronk2007,webber2011}.

This article asks a narrow but demanding question: in Bolivia, did economic liberalization reduce poverty and inequality, or did it increase them? The answer cannot be inferred from regional averages. In the Latin American panel, some indices of economic freedom may correlate with better average outcomes because they partly capture monetary stability, legal predictability, and reduced macroeconomic disorder. Bolivia is different because the concrete historical bundle of reform was implemented through severe adjustment, mining-sector collapse, relocalization, informality, capitalization, and export dependence. The empirical strategy therefore estimates a Bolivia-specific heterogeneous effect rather than assuming that the Latin American average applies to Bolivia.

The article contributes to three literatures. First, it contributes to Bolivian political economy by combining the historical interpretation of neoliberalism with a systematic reconstruction of macro, export, and labor evidence. Second, it contributes to the Latin American reform literature by showing that the average regional meaning of liberalization cannot simply be imposed on a historically distinctive case. Third, it contributes methodologically by combining a heterogeneous IV design with mechanism evidence drawn from export structure and harmonized historical household surveys.

\begin{findingbox}{Argument in one sentence}
Bolivian neoliberalism stabilized inflation, but its historically concrete liberalization package appears to have increased poverty, vulnerability, and inequality relative to the counterfactual implied by the heterogeneous regional-diffusion IV design.
\end{findingbox}

The main estimate is substantive rather than merely statistical. A 10-point increase in the Heritage economic-freedom score is associated, for Bolivia, with approximately +4.46 percentage points of poverty at the USD 4.20/day line, +3.61 percentage points at the USD 3/day line, +7.40 percentage points at the USD 8.30/day line, and +3.91 Gini points. These results are not presented as a universal law against markets. They are presented as evidence that Bolivia's actual neoliberal package produced a regressive social effect under the identification assumptions stated below.

\section{Historical sequence and political economy}

The Bolivian neoliberal cycle should be read as a sequence of regimes rather than as a single policy. The first regime was crisis and stabilization. The second was labor and state-enterprise restructuring, especially in mining. The third was second-generation institutional reform, including capitalization and decentralization. The fourth was the social conflict cycle of 2000--2005. The fifth was the post-neoliberal settlement, which did not abolish commodity dependence but redistributed rents and changed the political meaning of sovereignty.

\begin{figure}[!htbp]
\centering
\begin{adjustbox}{max width=\linewidth}
\begin{tikzpicture}[
  node distance=.8cm and .85cm,
  period/.style={rectangle, rounded corners, draw=blue!55!black, fill=blue!5, align=center, text width=2.8cm, minimum height=.85cm},
  arr/.style={-Latex, thick, blue!55!black}
]
\node[period] (a) {1982--1985\\monetary and fiscal collapse};
\node[period, right=of a] (b) {1985--1989\\DS 21060 and emergency stabilization};
\node[period, right=of b] (c) {1989--1993\\continuity, mining restructuring, labor displacement};
\node[period, below=1.0cm of c] (d) {1993--1997\\capitalization and Popular Participation};
\node[period, left=of d] (e) {1998--2005\\social conflict, water, gas, party-system crisis};
\node[period, left=of e] (f) {2006 onward\\post-neoliberal redistribution and extractive continuity};
\draw[arr] (a)--(b); \draw[arr] (b)--(c); \draw[arr] (c)--(d); \draw[arr] (d)--(e); \draw[arr] (e)--(f);
\end{tikzpicture}
\end{adjustbox}
\caption{Periodization of Bolivian neoliberalism. The sequence moves from emergency stabilization to social conflict and rent reconfiguration.}
\label{fig:timeline}
\end{figure}
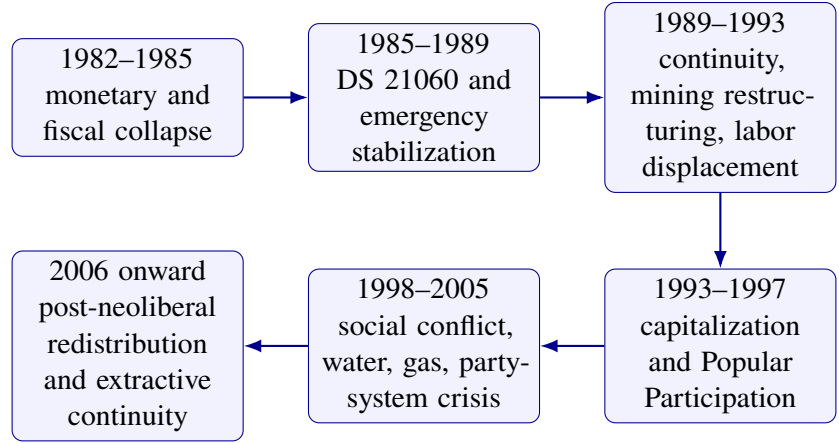

\input{tables/tbl_historical_mechanisms.tex}

The longer historical arc matters because Bolivian neoliberalism did not emerge in an institutional vacuum. The National Revolution of 1952 had built a state-centered model around nationalized mining, agrarian reform, organized labor, and a corporatist relation between state and popular sectors. That model was never internally homogeneous: it combined mass incorporation, regional inequalities, peasant differentiation, and dependence on mineral exports \cite{klein_bolivia,dunkerley1984,hylton_thomson2007}. By the late 1970s and early 1980s, the older developmental state faced a simultaneous crisis of external finance, fiscal capacity, mineral earnings, political authority, and monetary credibility. The neoliberal turn was therefore both a response to collapse and a project of state reconstruction.

The collapse of the tin-centered order was especially important. Tin had been more than an export commodity: it structured fiscal revenue, union power, regional settlement, and the political centrality of miners. The sharp deterioration of the mining economy in the mid-1980s weakened the material base of the old labor regime and made adjustment socially explosive. Relocalization should therefore be read not only as a labor-market event but as a transformation of Bolivian political society. It displaced households, reconfigured urban labor markets, strengthened informal and cooperative sectors, and contributed to the emergence of new territorial movements outside the old mining-union axis \cite{comibol,kohl_farthing2006,webber2011}.

The reforms of the 1990s changed the institutional form of the state rather than merely reducing it. Capitalization reorganized the relation between public assets, foreign investment, pension-linked shares, and sectoral regulation. Popular Participation decentralized resources and recognized local organizations, but it also helped stabilize a reform regime by relocating conflict and responsibility to municipalities \cite{ley1544,ley1551,kohl2002,barja2004}. The result was a paradox: a state that was smaller in some strategic enterprise functions but more territorially present in municipal politics. This is why the Bolivian case cannot be understood through the narrow formula of simple privatization. It was a reconfiguration of state capacity, social representation, and the distribution of rents.

The conflicts of 2000-2005 were consequently not isolated protests against tariffs or export routes. They condensed accumulated grievances around water, gas, indigenous citizenship, territorial autonomy, employment, and the legitimacy of the party system. The Guerra del Agua and the Guerra del Gas expressed a deeper contradiction: the neoliberal state could govern stabilization and contracts, but it could not produce a socially legitimate settlement over strategic resources and distribution \cite{perreault2006,spronk2007,webber2011}. The empirical analysis below should be read against that history: the coefficient on Bolivian liberalization is not a coefficient on an abstract market principle, but on a historically embedded reform bundle.

The 1985 stabilization must be understood against a genuine crisis. Sachs and Morales describe Bolivia's hyperinflation as one of the most dramatic episodes of the 1980s, while CEPAL interprets the post-1985 program as a combination of orthodox stabilization and structural reform \cite{sachs_morales,cepal_bolivia}. The monetary effect of the adjustment was not incidental. A society cannot maintain stable wages, tax collection, credit, or public planning when the unit of account collapses. This is why a serious critique of Bolivian neoliberalism should not deny stabilization. The sharper argument is that stabilization was converted into a broader social project that did not create inclusive development.

Mining is central to that social history. The restructuring of COMIBOL and the collapse of the old mining-labor order displaced thousands of workers and weakened one of the central pillars of twentieth-century Bolivian popular politics \cite{comibol,kohl_farthing2006,webber2011}. Relocalized workers moved into commerce, transport, informal urban occupations, cooperative mining, and coca-growing regions. The political consequence was not only unemployment. It was the reorganization of popular society itself. Organized mining labor lost centrality, but new urban, indigenous, neighborhood, peasant, cocalero, and territorial movements gained strategic importance.

The 1990s added institutional complexity. Capitalization was not a simple sale of public enterprises, but a hybrid arrangement in which strategic investors obtained control and investment obligations while shares were linked to pension-related distribution mechanisms \cite{ley1544,mesa_lago2002,barja2004}. Popular Participation decentralized parts of the state and transformed local politics, while also stabilizing the broader reform order \cite{ley1551,kohl2002}. These reforms coincided with the entrenchment of a liberalized policy regime, but they did not eliminate the structural limits of a highly unequal and commodity-dependent economy.

The social conflict cycle of the early 2000s revealed those limits. The water conflict in Cochabamba, the gas conflict, and the collapse of the party system were not accidental interruptions. They were part of the political economy of a regime that had stabilized money but not legitimacy \cite{perreault2006,spronk2007,webber2011}. When poverty fell after 2006, it did so under a different political settlement that combined commodity rents, redistribution, public investment, and a discursive reconfiguration of sovereignty. This later decline in poverty is therefore not evidence that the 1990s reform bundle was socially benign.

\section{Data architecture and descriptive evidence}

The empirical strategy combines several layers of evidence: annual macroeconomic and social indicators from the World Bank and related international databases; economic-freedom measures from Heritage and Fraser; labor-market and productivity indicators from ILOSTAT, PWT, and Maddison; IMF information from MONA and the Abiad--Detragiache--Tressel financial-reform database; disaster and fiscal-consolidation controls; annual Bolivian export series; and harmonized Bolivian historical surveys for labor-market mechanisms. The purpose of this architecture is not only scale. It is triangulation. The main IV estimate uses country-year panel information, the export data clarify an important channel, and the surveys show how macro restructuring translated into labor-market vulnerability.

\input{tables/tbl_coverage_clean.tex}

Table \ref{tab:coverage} clarifies the temporal support of the analysis. The main Bolivia poverty series begin in 1990, while Heritage begins in 1995. This means that the paper does not directly identify the immediate short-run social effect of DS 21060 in 1985. Instead, the preferred IV estimate identifies the effect of the liberalization regime during the years in which a comparable poverty or inequality outcome overlaps with the annual liberalization proxy. That temporal distinction matters and the interpretation throughout the paper respects it.

\begin{figure}[!htbp]
\centering
\includegraphics[width=.88\linewidth]{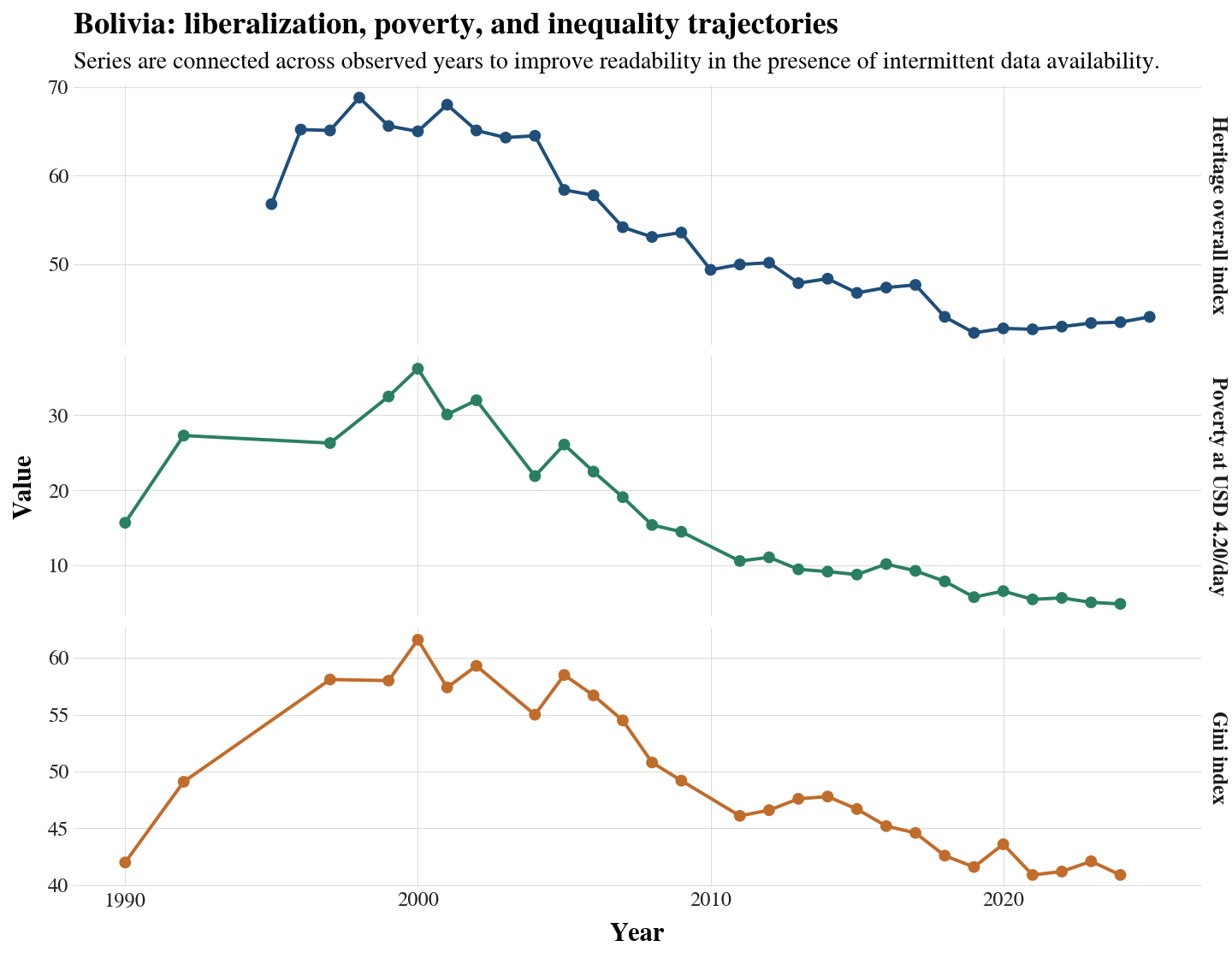}
\caption{Bolivia: trajectories of economic freedom, poverty at USD 4.20/day, and the Gini index. The lines connect observed data points across years in order to preserve the visual continuity of each series while leaving the substantive interpretation to the text and the econometric analysis.}
\label{fig:traj}
\end{figure}

Figure \ref{fig:traj} summarizes the core descriptive puzzle. High-Heritage years coincide with high poverty and high inequality in Bolivia, especially during the late 1990s and early 2000s. After 2006, poverty and inequality fall while Heritage declines. A purely historical reading therefore supports the intuition that the liberalization period coincided with severe social costs. However, historical alignment is not identical to causal inference. The post-2006 decline in poverty could reflect commodity rents, policy changes, and political regime shifts rather than the direct inverse of liberalization. This is why descriptive evidence must be complemented by an explicit identification strategy.

\input{tables/tbl_corr_clean.tex}

\begin{figure}[!htbp]
\centering
\includegraphics[width=.88\linewidth]{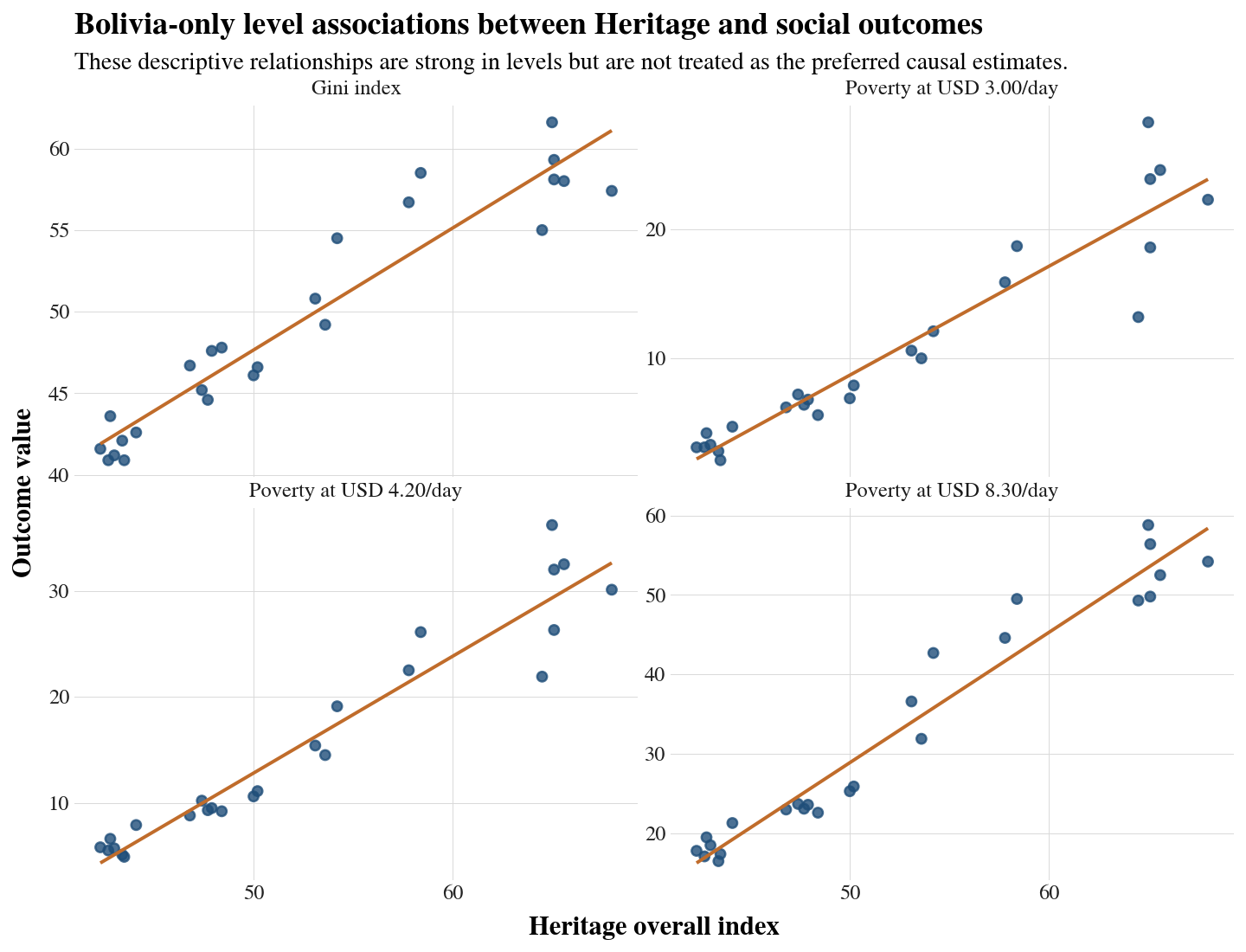}
\caption{Bolivia-only level associations between Heritage and social outcomes. These relationships are useful for descriptive orientation, but the article does not interpret them as the preferred causal estimates.}
\label{fig:corrfig}
\end{figure}

Table \ref{tab:corr} makes the descriptive point sharper. In levels, the Bolivia-only correlations between Heritage and poverty or inequality are strongly positive. In first differences, they become weak and unstable. Figure \ref{fig:corrfig} visualizes the same fact. The conclusion is methodological as much as descriptive: Bolivia-only levels evidence is highly suggestive historically, but it is not sufficient for causal inference because the series are trend-dominated and because regime changes create structural breaks.

\input{tables/tbl_ols_clean.tex}

The descriptive OLS estimates reinforce the same message. Table \ref{tab:ols} shows that a one-point increase in Heritage is associated with large increases in poverty and inequality in Bolivia-only levels regressions. Yet those estimates are too large and too fragile to be the main result. They are vulnerable to time trends, small-sample problems, and regime shifts. Their role is to document the historical alignment between liberalization and social deterioration, not to anchor the main causal claim.

\section{Identification strategy}

The preferred empirical design estimates a heterogeneous two-stage least-squares model in a Latin American panel and allows the Bolivia-specific treatment effect to differ from the regional average. The key idea is that regional reform waves may shift domestic liberalization through diffusion, imitation, conditionality, and technocratic learning. The instrument is the lagged leave-one-out regional average of Heritage:
\begin{equation}
Z_{it}=\frac{1}{N_{r}-1}\sum_{j\neq i, r}Heritage_{j,t-1}.
\end{equation}
To estimate a Bolivia-specific effect, the interacted excluded instrument is:
\begin{equation}
Z_{it}^{BOL}=Z_{it}\times \mathbf{1}\{i=Bolivia\}.
\end{equation}
The first-stage equations are:
\begin{align}
Heritage_{it} &= \pi_1 Z_{it}+\pi_2 Z_{it}^{BOL}+X'_{it}\delta+\mu_i+\lambda_t+u_{it},\\
Heritage_{it}\times Bolivia_i &= \rho_1 Z_{it}+\rho_2 Z_{it}^{BOL}+X'_{it}\theta+\mu_i+\lambda_t+v_{it}.
\end{align}
The second stage is:
\begin{equation}
Y_{it}=\beta_1\widehat{Heritage}_{it}+\beta_2\widehat{Heritage_{it}\times Bolivia_i}+X'_{it}\gamma+\mu_i+\lambda_t+\varepsilon_{it}.
\end{equation}
The Bolivia-specific effect is $\beta_{BOL}=\beta_1+\beta_2$, and the reported 10-point effect is $10\times\beta_{BOL}$.

The baseline controls are growth of GDP per capita, CPI inflation, and the terms of trade. These controls matter because monetary normalization, business-cycle fluctuations, and external-price shocks are all plausible confounders. The export-controls specification adds Bolivian export shares and concentration measures. That model is intentionally interpreted as a mechanism and sensitivity check rather than as a more definitive causal specification, because export structure may itself lie on the causal path from liberalization to social outcomes.

\begin{figure}[!htbp]
\centering
\begin{adjustbox}{max width=.92\linewidth}
\begin{tikzpicture}[
  node distance=.9cm and 1.1cm,
  znode/.style={rectangle, rounded corners, draw=green!45!black, fill=green!6, align=center, text width=3.4cm, minimum height=.9cm},
  tnode/.style={rectangle, rounded corners, draw=blue!50!black, fill=blue!6, align=center, text width=3.4cm, minimum height=.9cm},
  ynode/.style={rectangle, rounded corners, draw=red!50!black, fill=red!6, align=center, text width=3.4cm, minimum height=.9cm},
  cnode/.style={rectangle, rounded corners, draw=black!45, fill=gray!8, align=center, text width=3.2cm, minimum height=.85cm},
  arr/.style={-Latex, thick}
]
\node[znode] (z) {Lagged regional diffusion\\leave-one-out Heritage};
\node[tnode, right=of z] (t) {Domestic liberalization\\Heritage};
\node[ynode, right=of t] (y) {Poverty and inequality};
\node[cnode, below=of t] (x) {Growth, inflation, terms of trade, export controls};
\node[cnode, above=of t] (fe) {Country and year fixed effects};
\draw[arr] (z)--(t);
\draw[arr] (t)--(y);
\draw[arr] (x)--(t); \draw[arr] (x)--(y);
\draw[arr] (fe)--(t); \draw[arr] (fe)--(y);
\end{tikzpicture}
\end{adjustbox}
\caption{Instrumental-variables design. Identification requires that lagged regional diffusion affects Bolivian poverty and inequality through domestic liberalization after controls and fixed effects.}
\label{fig:ivdiagram}
\end{figure}
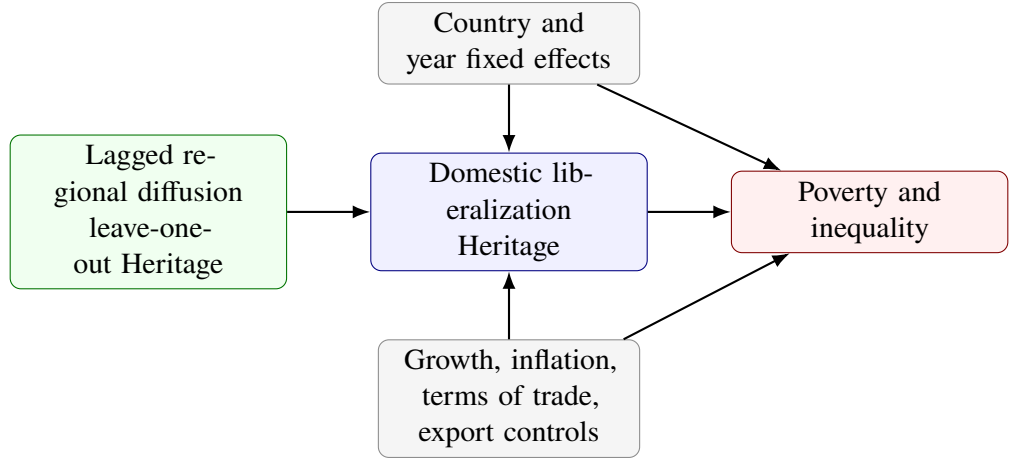

\input{tables/tbl_first_stage_clean.tex}

\begin{figure}[!htbp]
\centering
\includegraphics[width=.84\linewidth]{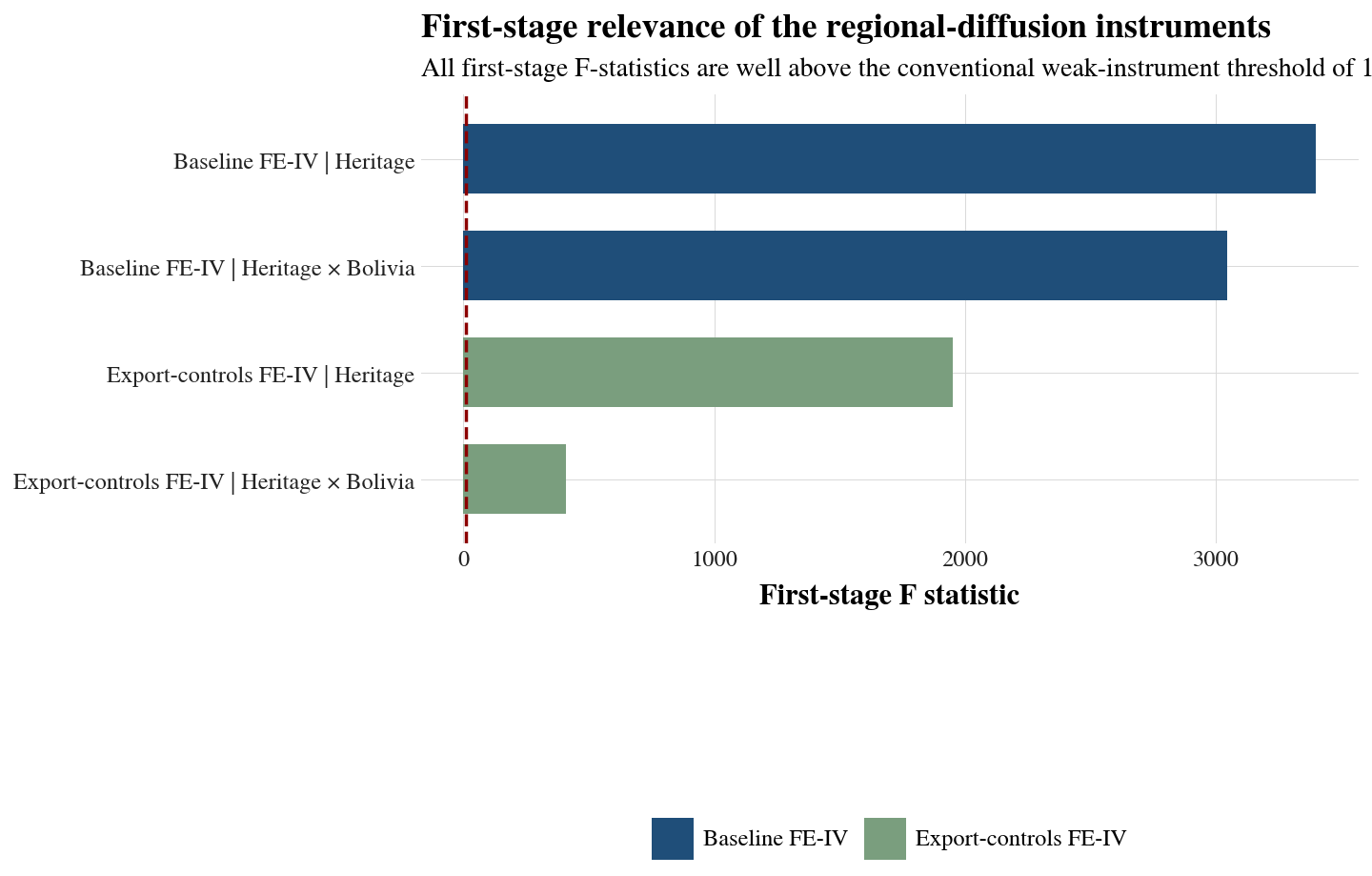}
\caption{First-stage relevance of the regional-diffusion instruments. The dashed line marks the conventional F=10 threshold.}
\label{fig:firststage}
\end{figure}

The first-stage results are decisive on relevance. Table \ref{tab:firststage} and Figure \ref{fig:firststage} show F statistics far above the conventional weak-instrument threshold of 10. The baseline specification exceeds 700 for both endogenous terms, and the export-controls specification remains strong. The empirical risk is therefore not weak instrumentation. The real risk is exclusion. Regional policy diffusion may be correlated with IMF conditionality, regional macro shocks, debt cycles, or political alignment. The paper addresses this risk with country and year fixed effects and a set of macro controls, but the exclusion restriction remains an identifying assumption rather than a demonstrable fact.

\section{Main empirical results}

\input{tables/tbl_iv_baseline_clean.tex}

Table \ref{tab:ivbaseline} contains the core results of the paper. The average regional coefficient is positive but modest. The crucial quantity is the Bolivia-specific total, which adds the Bolivia interaction to the regional coefficient. For Bolivia, a 10-point increase in the Heritage score is associated with +4.46 percentage points of poverty at USD 4.20/day, +3.61 percentage points at USD 3/day, +7.40 percentage points at USD 8.30/day, and +3.91 Gini points. These are economically large effects. They imply not only more poverty among the poor, but also a widening of vulnerability around a broader lower-middle-income threshold and a rise in inequality.

The pattern across outcomes is substantively informative. The largest poverty effect appears at USD 8.30/day, suggesting that the Bolivia-specific liberalization effect was especially strong in the broader zone of vulnerability rather than only at the extreme-poverty margin. This is historically plausible. Labor displacement, weak formal absorption, and uneven access to new opportunities are more likely to enlarge the vulnerable population than to operate only at the very bottom of the distribution. The positive Gini effect is also large enough to matter politically. A nearly four-point Gini increase for a 10-point Heritage shift is fully consistent with a social regime in which gains from liberalization were distributed asymmetrically.

\begin{resultbox}{Bolivia-specific result}
Under the heterogeneous regional-diffusion IV design, Bolivia's liberalization effect is positive for poverty and inequality: a 10-point Heritage increase implies approximately $+4.46\pp$ poverty at USD 4.20/day, $+3.61\pp$ poverty at USD 3/day, $+7.40\pp$ poverty at USD 8.30/day, and $+3.91$ Gini points.
\end{resultbox}

\begin{figure}[!htbp]
\centering
\includegraphics[width=.84\linewidth]{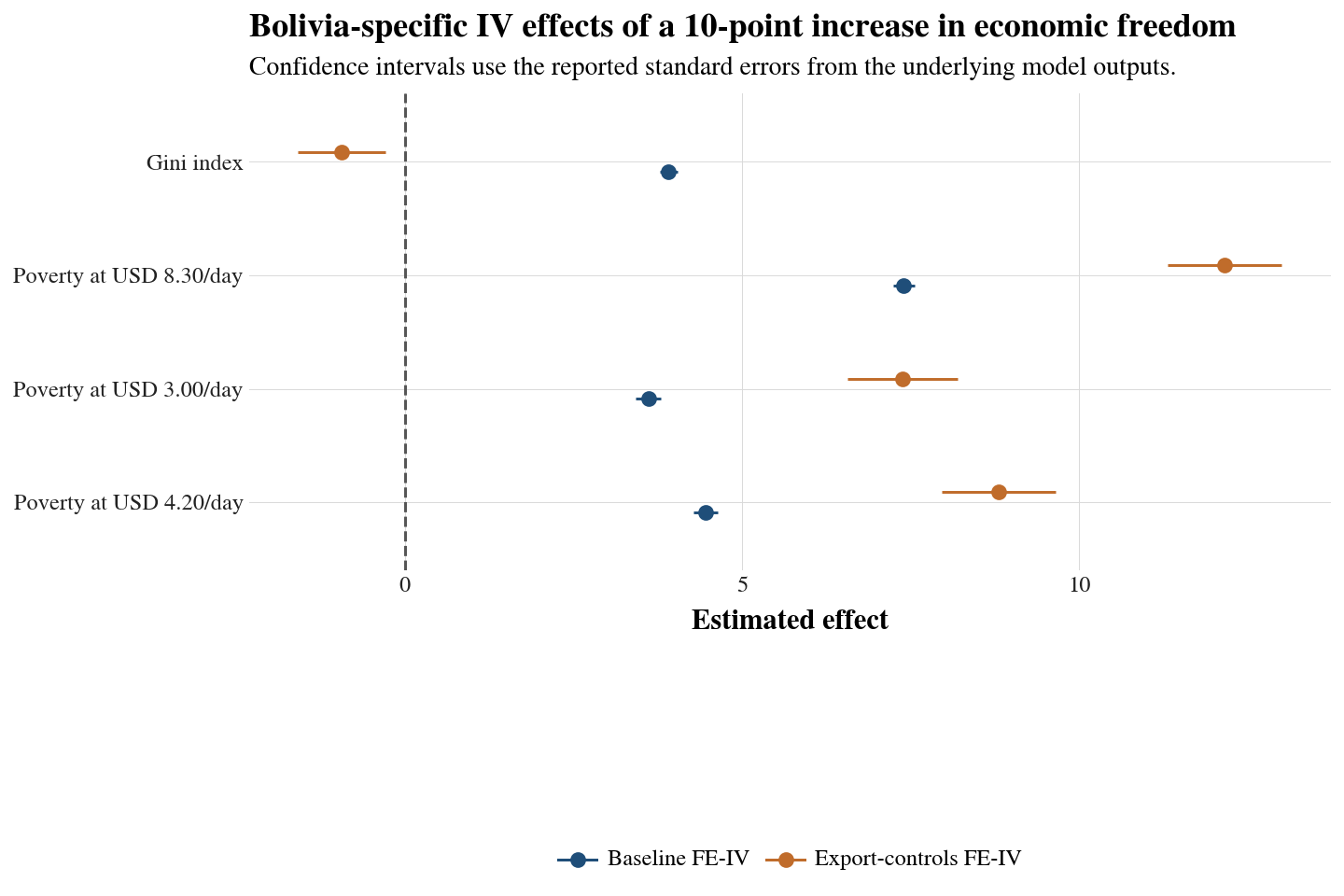}
\caption{Bolivia-specific heterogeneous IV effects of a 10-point Heritage increase. Baseline and export-controls specifications are shown together for direct comparison.}
\label{fig:iveffects}
\end{figure}

The IV result should not be interpreted as a universal anti-market theorem. It is a Bolivia-specific estimate under assumptions. In many countries, a market-reform index may partly capture legal stability, monetary normalization, and reduced macroeconomic disorder. In Bolivia, however, the same index overlaps with a historically distinct bundle of labor displacement, mining restructuring, privatization/capitalization, and weak formal-employment absorption. The role of the Bolivia interaction is precisely to prevent the regional average from erasing that specific historical trajectory.

\section{Export dependence and the limits of market-led development}

Bolivia's export structure helps explain how liberalization and macroeconomic stabilization could coexist with social vulnerability. An externally oriented commodity economy can generate foreign exchange and fiscal revenue without creating broad-based formal employment. The export data show the long-run importance of hydrocarbons and minerals, the rise of the gas cycle in the 2000s, and persistent concentration by product and destination.

\begin{figure}[!htbp]
\centering
\includegraphics[width=.86\linewidth]{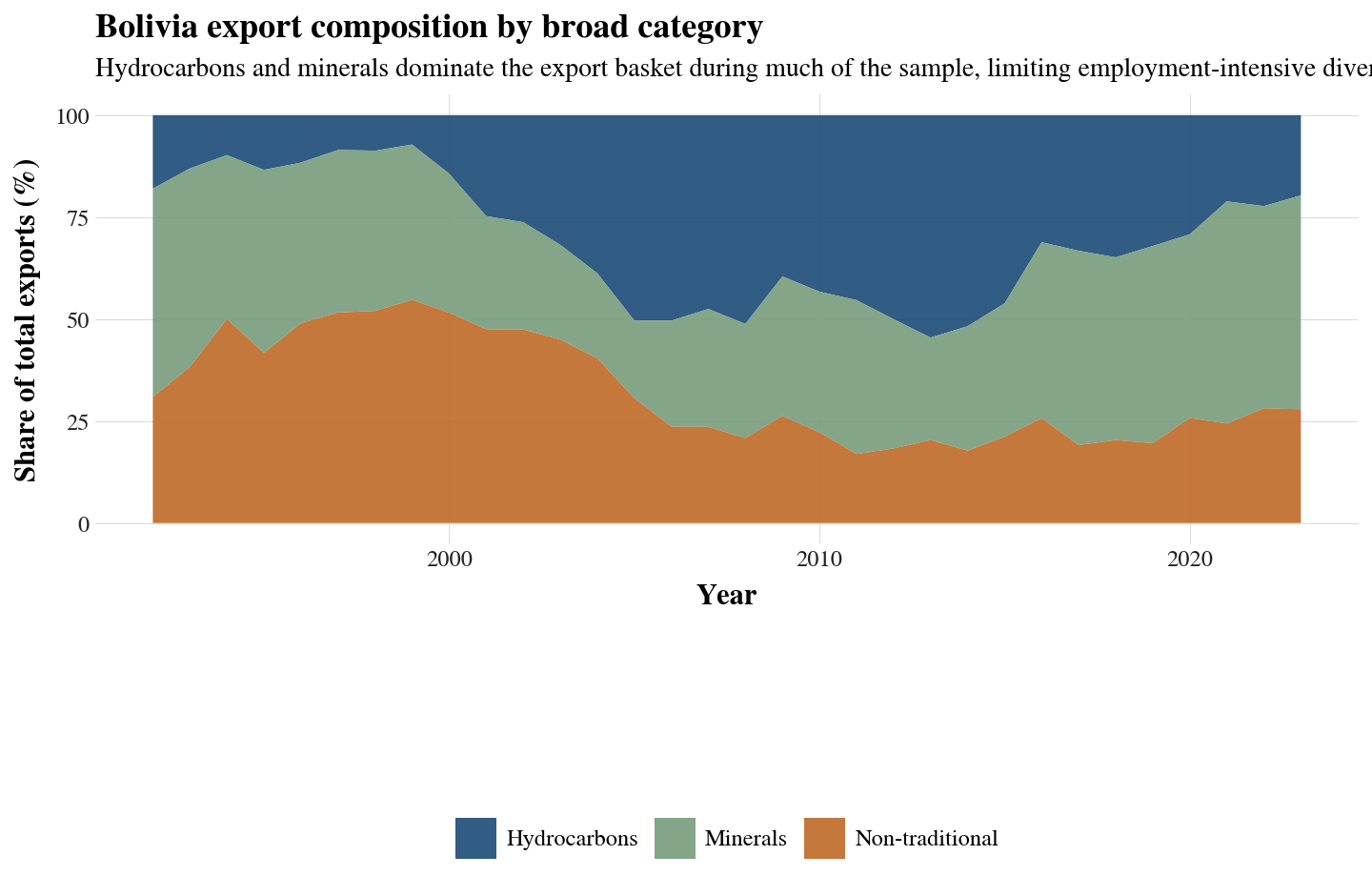}
\caption{Bolivia export composition by broad category. The dominance of hydrocarbons and minerals helps explain why growth and foreign-exchange gains did not automatically translate into inclusive labor-market transformation.}
\label{fig:exportscomp}
\end{figure}

\begin{figure}[!htbp]
\centering
\includegraphics[width=.86\linewidth]{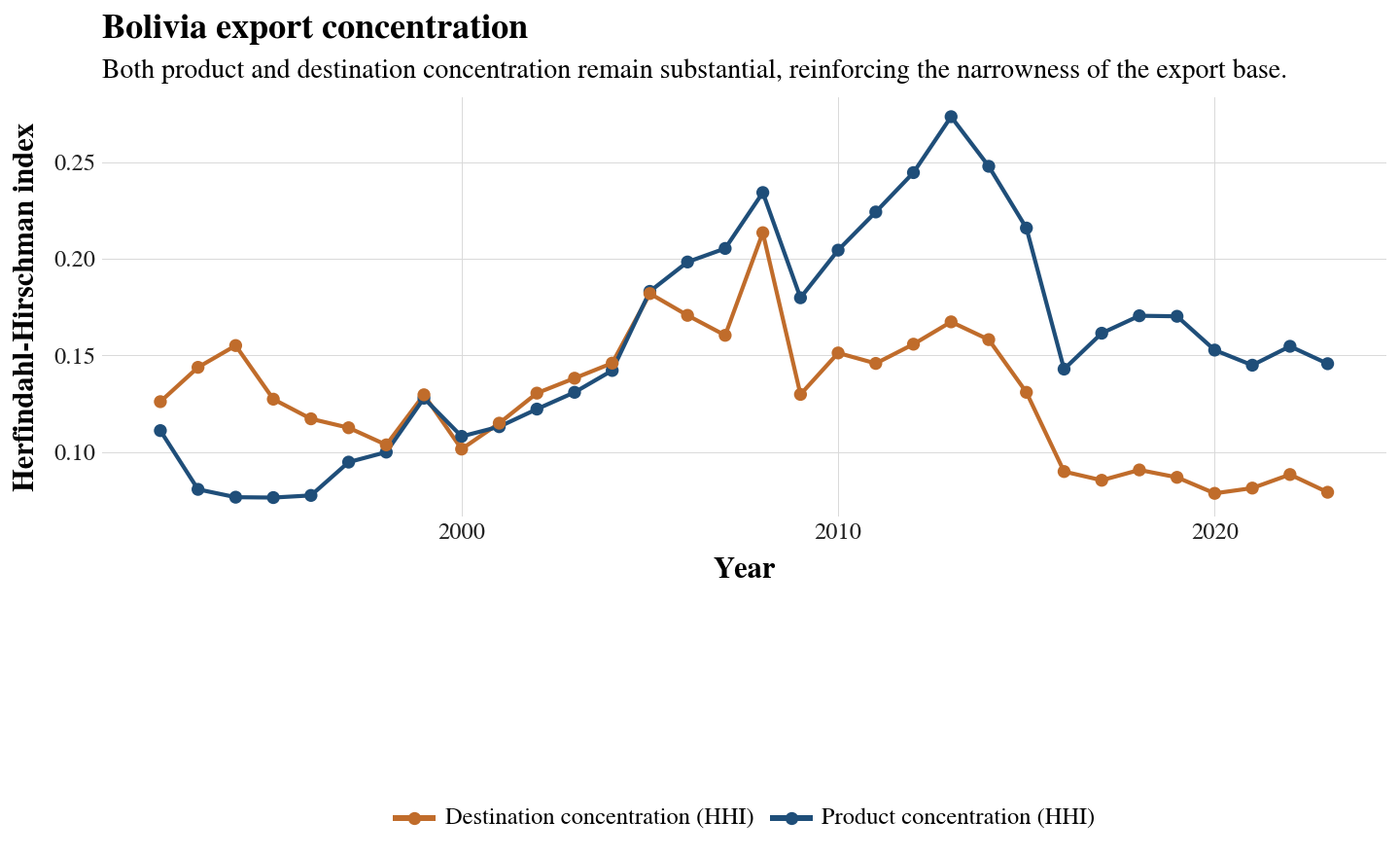}
\caption{Bolivia export concentration. Persistent concentration by product and destination reveals the narrowness of the export base.}
\label{fig:exporthhi}
\end{figure}

\input{tables/tbl_exports_clean.tex}

Table \ref{tab:exports} converts the visual impression into concrete magnitudes. Traditional exports remain dominant across the sample, and by 2005 and 2014 hydrocarbons alone accounted for more than 50\% of exports. Product concentration rises sharply during the commodity boom. Destination concentration is also substantial, although somewhat lower in recent years. These numbers matter because they identify the type of outward orientation embedded in the Bolivian liberalization regime: not a diversified export transformation with broad labor absorption, but a concentrated commodity structure with strong rent dependence.

\input{tables/tbl_iv_export_clean.tex}

The export-controls specification deepens the interpretation. Table \ref{tab:ivexport} shows that the sign of the poverty effect remains positive when export-structure controls are added, although precision deteriorates for some outcomes and the Gini estimate becomes imprecise. This is exactly what one would expect if export structure is part confounder and part mechanism. Controlling it may reduce omitted-variable bias, but it may also partial out part of the causal channel through which liberalization affects social outcomes. The export-controls model should therefore be read as a disciplined sensitivity exercise rather than a replacement for the baseline specification.

The post-2006 decline in poverty can also be clarified in this framework. Poverty fell after the neoliberal period even as Heritage scores declined. That does not mean that lower economic freedom mechanically reduced poverty. It means that the social regime changed. Hydrocarbon rents, redistribution, public investment, and a different political settlement altered how external income was collected and distributed. The fall in poverty after 2006 is therefore partly a rent-distribution and political-economy story rather than the simple mirror image of the 1990s reform sequence.

\section{Labor mechanisms from historical surveys}

The harmonized historical surveys provide mechanism evidence for 1989--2005. These surveys do not reconstruct a fully comparable official poverty line across the whole period, so they are not used as the main poverty outcome. Their value lies elsewhere: they allow the analysis to examine employment, unemployment, informality proxies, labor income, and income dispersion. Labor mechanisms matter because the social effect of neoliberalism should not be evaluated only through aggregate GDP or inflation.

\input{tables/tbl_surveys_clean.tex}

Table \ref{tab:surveys} suggests several features of the labor regime. Employment rates do not collapse across the board, but that is not a sign of social success. In a low-income and highly informal economy, workers cannot afford to remain openly unemployed. Distress is absorbed through self-employment, precarious informal work, underemployment, and low earnings. The relatively high informality proxy in 1986--1990 and 2001--2005, together with the still-high labor-income inequality, fits the argument that the old labor order was dismantled without a socially inclusive formal alternative. The decline in the labor-income Gini by 2001--2005 is informative, but it should not be over-read as an egalitarian success: falling labor-income inequality can coexist with widespread low-productivity informal absorption.

\begin{figure}[!htbp]
\centering
\includegraphics[width=.86\linewidth]{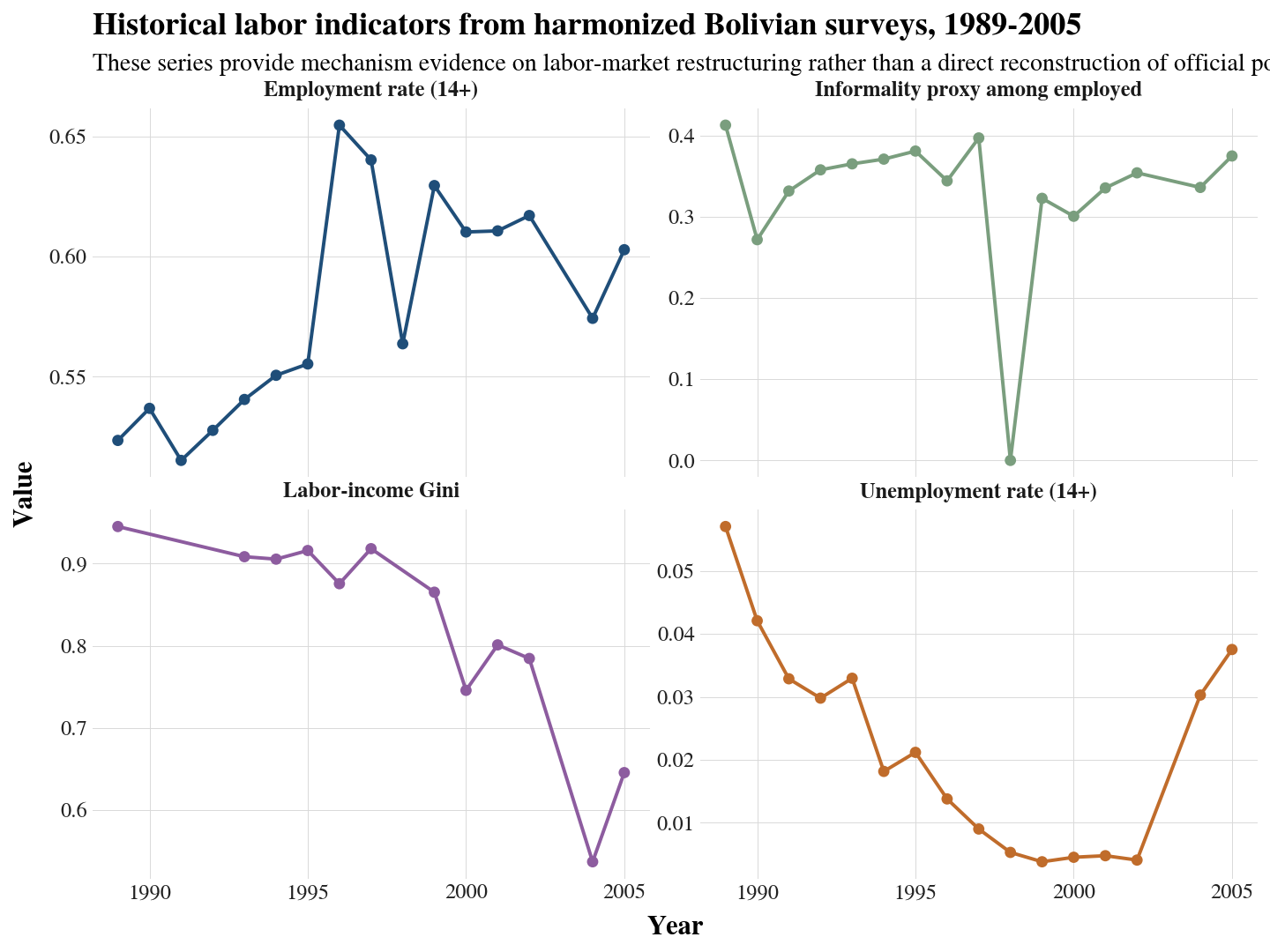}
\caption{Historical labor indicators from harmonized Bolivian surveys, 1989--2005. These series support the labor-mechanism interpretation rather than serving as official poverty estimates.}
\label{fig:surveys}
\end{figure}

The surveys therefore connect the macro and social layers of the argument. Stabilization solved one type of crisis---the destruction of money---but the labor market absorbed adjustment through channels that are structurally compatible with poverty and vulnerability. This is why a purely macroeconomic defense of the period is insufficient. The political economy of employment and informal absorption matters for the total social effect.

\section{Diagnostics and robustness}

The empirical claim is strongest when stated with explicit assumptions. The main estimate is not a simple correlation. It is a heterogeneous IV result with strong first stages. Yet the model must confront heteroskedasticity, serial correlation, non-stationarity, and exclusion concerns. The diagnostics therefore play an interpretive role: they explain why the paper relies on robust inference and why the preferred design is a panel IV rather than a Bolivia-only levels regression.

\input{tables/tbl_diagnostics_clean.tex}

Table \ref{tab:diagnostics} shows that robust inference is necessary. Durbin--Watson statistics below two indicate positive serial correlation, and Breusch--Pagan p-values indicate heteroskedasticity. For that reason the article emphasizes robust standard errors and triangulation across sign, magnitude, historical consistency, and mechanism evidence rather than a single classical inference calculation. The positive sign of the Bolivia-specific effect is not a fragile artifact of one equation; it is consistent across outcomes and coherent with the broader historical narrative.

\input{tables/tbl_vif_clean.tex}

Variance inflation factors in Table \ref{tab:vif} are low in the baseline controls before fixed effects. This does not eliminate all concerns about collinearity, especially after adding fixed effects and interactions, but it does show that the core controls are not simply restating each other. In practical terms, the estimated social effect is not being driven by a trivial multicollinearity problem among growth, inflation, and the terms of trade.

\input{tables/tbl_adf_clean.tex}

The stationarity tests in Table \ref{tab:adf} reinforce the caution about Bolivia-only time-series models. Several poverty and inequality series are non-stationary in levels. This is why levels correlations in Bolivia, although historically meaningful, are not by themselves sufficient for causal inference. The panel IV design is preferred because it uses both temporal and cross-country variation while still retaining a Bolivia-specific coefficient.

\section{Causal interpretation}

The empirical evidence supports a layered interpretation. First, the stabilization component of Bolivian neoliberalism reduced hyperinflation and restored monetary order. Second, the social-restructuring component increased vulnerability through labor displacement, informality, and unequal access to the gains from liberalization. Third, the export structure channeled the gains of outward orientation through commodities and rents rather than through broad formal employment. Fourth, the later reduction in poverty reflected a changed regime of rent distribution and political incorporation rather than a straightforward continuation of the 1990s policy logic.

The causal object is therefore not an abstract market principle. It is the historical bundle actually implemented in Bolivia. Formally, the observed treatment can be written as:
\begin{align}
Liberalization_{BOL,t}={}& MonetaryStabilization_t+LaborRestructuring_t+Capitalization_t \\
&+TradeOpening_t+FinancialReform_t+ExportDependence_t .
\end{align}
The social effect is the sum of countervailing channels:
\begin{align}
\frac{\partial Poverty}{\partial Liberalization}={}&
\underbrace{\frac{\partial Poverty}{\partial Stabilization}}_{<0}
+\underbrace{\frac{\partial Poverty}{\partial LaborDisplacement}}_{>0} \\
&+\underbrace{\frac{\partial Poverty}{\partial UnequalRents}}_{>0}
+\underbrace{\frac{\partial Poverty}{\partial FormalInvestment}}_{\leq 0\;?}.
\end{align}
The empirical estimate suggests that the poverty-increasing channels dominated the poverty-reducing channels for Bolivia.

\begin{figure}[!htbp]
\centering
\begin{adjustbox}{max width=.90\linewidth}
\begin{tikzpicture}[
  node distance=.75cm and .8cm,
  root/.style={rectangle, rounded corners, draw=black!65, fill=gray!8, align=center, text width=4.3cm, minimum height=.9cm},
  good/.style={rectangle, rounded corners, draw=green!55!black, fill=green!6, align=center, text width=3.5cm, minimum height=.85cm},
  neg/.style={rectangle, rounded corners, draw=red!55!black, fill=red!6, align=center, text width=3.5cm, minimum height=.85cm},
  outcomeBox/.style={rectangle, rounded corners, draw=blue!55!black, fill=blue!6, align=center, text width=3.6cm, minimum height=.85cm},
  arr/.style={-Latex, thick}
]
\node[root] (lib) {Bolivian neoliberal liberalization package};
\node[good, below left=1.0cm and .5cm of lib] (stab) {Monetary stabilization\\inflation relief};
\node[neg, below=1.0cm of lib] (labor) {Labor restructuring\\relocalization and informality};
\node[neg, below right=1.0cm and .5cm of lib] (export) {Commodity export dependence\\concentrated rents};
\node[outcomeBox, below=1.0cm of labor] (y) {Net social result\\higher poverty and inequality};
\draw[arr] (lib)--(stab); \draw[arr] (lib)--(labor); \draw[arr] (lib)--(export);
\draw[arr] (stab)--(y); \draw[arr] (labor)--(y); \draw[arr] (export)--(y);
\end{tikzpicture}
\end{adjustbox}
\caption{Mechanism map. The net estimated effect is positive for poverty because the poverty-increasing channels dominate the monetary-stabilization channel within Bolivia's historically specific liberalization package.}
\label{fig:mechanism}
\end{figure}
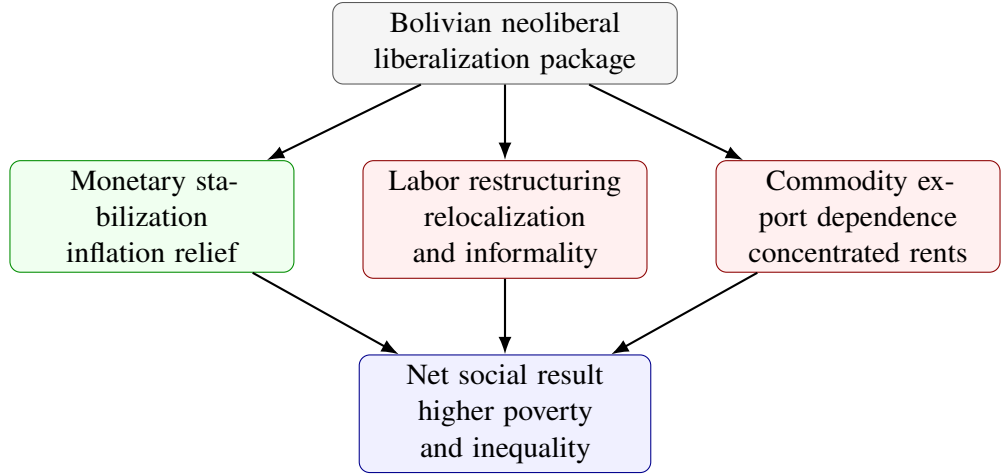

\section{Limitations}

The main limitation is the exclusion restriction. Regional policy diffusion is a plausible instrument because reform waves spread through technocratic networks, conditionality, imitation, and policy learning. Yet regional diffusion may also proxy for common shocks. The article controls for year effects, country effects, growth, inflation, terms of trade, and export structure, but no observational IV design can prove exclusion. A second limitation is measurement. Heritage is an aggregate index. It is useful for annual coverage but conceptually broad. Future work should disaggregate monetary freedom, trade freedom, fiscal burden, labor flexibility, investment openness, and financial liberalization. A third limitation is temporal. The main poverty-Heritage overlap begins after the initial 1985 shock. Therefore, the IV estimate captures the social consequences of the liberalization regime during the available overlap, not the entire immediate effect of DS 21060. A fourth limitation is survey comparability. The historical surveys are valuable for mechanisms but cannot replace a harmonized official poverty series.

\section{Conclusion}

Bolivia's neoliberal history cannot be reduced to a binary judgment. It was neither pure technocratic success nor pure social catastrophe. It was a historically specific sequence in which monetary stabilization succeeded while inclusive development failed. The empirical evidence developed here indicates that, for Bolivia, economic liberalization measured through Heritage is associated with higher poverty and higher inequality under a heterogeneous regional-diffusion IV design. The most policy-relevant headline estimate is:
\begin{equation}
+10\;Heritage\;points\Rightarrow +4.46\pp\;poverty_{USD4.20/day}+3.91\;Gini\;points.
\end{equation}
This result should be read as a Bolivia-specific causal estimate under assumptions. It does not imply that every market reform everywhere increases poverty. It implies that the Bolivian neoliberal package, embedded in mining restructuring, labor displacement, capitalization, informality, and commodity dependence, generated a regressive social outcome even while stabilizing the monetary system. More broadly, the article suggests that the political economy of liberalization matters as much as the abstract policy label. The social meaning of market-oriented reform depends on the production structure, the labor market, the state, and the distributional regime through which reform is enacted. In Bolivia, those conditions made stabilization possible, but they did not make liberalization socially inclusive.

The central historical lesson is therefore double. First, a progressive reading of Bolivian history should not minimize the destructive force of hyperinflation or the importance of restoring monetary order. Second, a technocratic reading should not confuse stabilization with development. The Bolivian experience shows that reforms can solve a nominal crisis while deepening a social crisis if labor absorption, productive diversification, and distributive legitimacy are not built into the policy regime. The empirical finding that Bolivia-specific liberalization increased poverty and inequality is consistent with that layered interpretation: the problem was not simply that markets expanded, but that the concrete expansion occurred through displacement, concentrated rents, weak formal employment creation, and a narrow export structure.

\FloatBarrier

\end{document}

%% file: tables/tbl_historical_mechanisms.tex
\begin{table}[!htbp]
\centering
\scriptsize
\caption{Historical mechanisms linking Bolivian neoliberalism to social outcomes}
\label{tab:historicalmechanisms}
\begin{adjustbox}{max width=\linewidth}
\begin{tabularx}{\linewidth}{p{2.2cm}p{3.1cm}p{3.0cm}p{3.5cm}X}
\toprule
Period & Policy-political core & Economic channel & Social channel & Interpretation \\
\midrule
1982--1985 & Fiscal, monetary, and external crisis & Hyperinflation, exchange-rate disorder, fiscal collapse & Real-wage destruction, uncertainty, erosion of state capacity & Stabilization demand was grounded in a real crisis rather than only ideology. \\
1985--1989 & DS 21060 and New Economic Policy & Price liberalization, fiscal compression, exchange-rate stabilization & Relocation, unemployment risk, informal absorption & Monetary stabilization reduced nominal chaos but shifted adjustment costs to labor. \\
1989--1993 & Continuity and mining restructuring & Decline of state mining, constrained public employment & Weakening of the mining-union axis, migration, cooperative and informal work & The old popular-corporatist settlement was dismantled without a broad formal-employment substitute. \\
1993--1997 & Capitalization and Popular Participation & Strategic investment, regulatory restructuring, municipal redistribution & New local political arenas, persistent inequality, partial access gains & The state was not simply withdrawn; it was reorganized territorially and sectorally. \\
1998--2005 & Water, gas, and legitimacy crisis & Commodity and service conflicts, export-route disputes & Neighborhood, indigenous, cocalero, and popular mobilization & Social conflict revealed the weak legitimacy of market-oriented control over strategic resources. \\
2006 onward & Post-neoliberal rent redistribution & Hydrocarbon rents, public investment, wage and transfer policy & Poverty reduction, new inclusion, extractive continuity & Poverty fell under a different distributive regime, not because the neoliberal sequence itself became inclusive. \\
\bottomrule
\end{tabularx}
\end{adjustbox}
\end{table}

%% file: tables/tbl_coverage_clean.tex
\begin{table}[!htbp]
\centering
\small
\caption{Bolivia variable coverage in the analytical dataset}
\label{tab:coverage}
\begin{adjustbox}{max width=\linewidth}
\begin{tabular}{lrrr}
\toprule
Variable & Years & First & Last \\
\midrule
Heritage economic freedom & 31 & 1995 & 2025 \\
Fraser EFW index & 30 & 1970 & 2023 \\
Poverty, USD 4.20/day & 27 & 1990 & 2024 \\
Poverty, USD 3/day & 27 & 1990 & 2024 \\
Poverty, USD 8.30/day & 27 & 1990 & 2024 \\
Gini index & 27 & 1990 & 2024 \\
Goods exports, million USD & 32 & 1992 & 2023 \\
Traditional export share & 32 & 1992 & 2023 \\
Hydrocarbons export share & 32 & 1992 & 2023 \\
Survey informality proxy & 16 & 1989 & 2005 \\
Survey labor-income Gini & 12 & 1989 & 2005 \\
\bottomrule
\end{tabular}
\end{adjustbox}
\end{table}

%% file: tables/tbl_corr_clean.tex
\begin{table}[!htbp]
\centering
\small
\caption{Bolivia-only correlations between Heritage and poverty or inequality}
\label{tab:corr}
\begin{adjustbox}{max width=\linewidth}
\begin{tabular}{lrrrr}
\toprule
Outcome & N levels & Level r & N diff. & Difference r \\
\midrule
Poverty USD 4.20/day & 25 & 0.963 & 24 & -0.103 \\
Poverty USD 3/day & 25 & 0.945 & 24 & -0.161 \\
Poverty USD 8.30/day & 25 & 0.974 & 24 & 0.049 \\
Gini & 25 & 0.953 & 24 & -0.221 \\
\bottomrule
\end{tabular}
\end{adjustbox}
\vspace{1mm}\begin{minipage}{0.95\linewidth}\footnotesize The level correlations are strongly positive, while first-difference correlations are weak. This is why the article does not rely on simple Bolivia-only correlations as causal evidence.\end{minipage}
\end{table}

%% file: tables/tbl_ols_clean.tex
\begin{table}[!htbp]
\centering
\scriptsize
\caption{Bolivia-only descriptive OLS estimates}
\label{tab:ols}
\begin{adjustbox}{max width=\linewidth}
\begin{tabular}{lrrrrrr}
\toprule
Outcome & N & Coef. per 1 & SE & p & Effect +10 & R2 \\
\midrule
Poverty USD 4.20/day & 24 & 1.091 & 0.092 & <0.001 & 10.91 & 0.936 \\
Poverty USD 3/day & 24 & 0.834 & 0.091 & <0.001 & 8.34 & 0.902 \\
Poverty USD 8.30/day & 24 & 1.613 & 0.089 & <0.001 & 16.13 & 0.957 \\
Gini & 24 & 0.727 & 0.058 & <0.001 & 7.27 & 0.918 \\
\bottomrule
\end{tabular}
\end{adjustbox}
\vspace{1mm}\begin{minipage}{0.95\linewidth}\footnotesize These are descriptive levels regressions, not the preferred causal estimates. The small Bolivia-only sample is vulnerable to time trends and regime shifts.\end{minipage}
\end{table}

%% file: tables/tbl_first_stage_clean.tex
\begin{table}[!htbp]
\centering
\scriptsize
\caption{First-stage relevance tests for the regional diffusion instruments}
\label{tab:firststage}
\begin{adjustbox}{max width=\linewidth}
\begin{tabular}{llrrrrr}
\toprule
Spec & Endogenous term & N & Countries & Years & First-stage F & R2 \\
\midrule
Baseline FE-IV & Heritage & 357 & 22 & 28 & 849.2 & 0.832 \\
Baseline FE-IV & Heritage x Bolivia & 357 & 22 & 28 & 761.2 & 0.814 \\
Export controls FE-IV & Heritage & 357 & 22 & 28 & 487.7 & 0.834 \\
Export controls FE-IV & Heritage x Bolivia & 357 & 22 & 28 & 101.9 & 0.959 \\
\bottomrule
\end{tabular}
\end{adjustbox}
\end{table}

%% file: tables/tbl_iv_baseline_clean.tex
\begin{table}[!htbp]
\centering
\scriptsize
\caption{Baseline heterogeneous IV estimates. Effects are measured per one Heritage point except the last column.}
\label{tab:ivbaseline}
\begin{adjustbox}{max width=\linewidth}
\begin{tabular}{lrrrrrrrr}
\toprule
Outcome & N & Countries & Years & Regional & Bolivia total & SE & p & Effect +10 \\
\midrule
Poverty USD 4.20/day & 357 & 22 & 28 & 0.111 & 0.446 & 0.094 & <0.001 & 4.46 \\
Poverty USD 3/day & 357 & 22 & 28 & 0.051 & 0.361 & 0.096 & <0.001 & 3.61 \\
Poverty USD 8.30/day & 357 & 22 & 28 & 0.138 & 0.740 & 0.080 & <0.001 & 7.40 \\
Gini & 357 & 22 & 28 & 0.024 & 0.391 & 0.067 & <0.001 & 3.91 \\
\bottomrule
\end{tabular}
\end{adjustbox}
\vspace{1mm}\begin{minipage}{0.95\linewidth}\footnotesize Regional is the average regional coefficient. Bolivia total is the regional coefficient plus the Bolivia-specific interaction. Effect +10 reports the implied Bolivia effect of a 10-point Heritage increase.\end{minipage}
\end{table}

%% file: tables/tbl_exports_clean.tex
\begin{table}[!htbp]
\centering
\scriptsize
\caption{Selected years of Bolivia export structure, 1992-2023}
\label{tab:exports}
\begin{adjustbox}{max width=\linewidth}
\begin{tabular}{rrrrrrrr}
\toprule
Year & Total exports & Traditional \% & Hydrocarbons \% & Minerals \% & Non-traditional \% & Product HHI & Destination HHI \\
\midrule
1992 & 773.8 & 69.1 & 18.0 & 51.1 & 30.9 & 0.111 & 0.126 \\
1997 & 1,272.1 & 48.4 & 8.5 & 39.8 & 51.6 & 0.095 & 0.113 \\
2000 & 1,475.0 & 48.4 & 14.3 & 34.1 & 51.6 & 0.108 & 0.102 \\
2005 & 2,948.1 & 69.4 & 50.3 & 19.0 & 30.6 & 0.183 & 0.182 \\
2010 & 7,052.1 & 77.7 & 43.3 & 34.5 & 22.3 & 0.204 & 0.151 \\
2014 & 13,034.2 & 82.2 & 51.7 & 30.5 & 17.8 & 0.248 & 0.158 \\
2019 & 8,933.3 & 80.4 & 32.1 & 48.3 & 19.6 & 0.170 & 0.087 \\
2023 & 10,935.0 & 72.1 & 19.7 & 52.4 & 27.9 & 0.146 & 0.079 \\
\bottomrule
\end{tabular}
\end{adjustbox}
\end{table}

%% file: tables/tbl_iv_export_clean.tex
\begin{table}[!htbp]
\centering
\small
\caption{Heterogeneous IV estimates with Bolivia export-structure controls}
\label{tab:ivexport}
\begin{adjustbox}{max width=\linewidth}
\begin{tabular}{lrrrr}
\toprule
Outcome & Effect +10 & SE & p & R2 \\
\midrule
Poverty USD 4.20/day & 8.81 & 0.430 & 0.041 & 0.168 \\
Poverty USD 3/day & 7.38 & 0.416 & 0.077 & 0.117 \\
Poverty USD 8.30/day & 12.15 & 0.432 & 0.005 & 0.205 \\
Gini & -0.94 & 0.330 & 0.776 & 0.168 \\
\bottomrule
\end{tabular}
\end{adjustbox}
\vspace{1mm}\begin{minipage}{0.95\linewidth}\footnotesize Export controls are interpreted as mechanism and sensitivity controls because export structure may itself lie on the causal path from liberalization to poverty.\end{minipage}
\end{table}

%% file: tables/tbl_surveys_clean.tex
\begin{table}[!htbp]
\centering
\scriptsize
\caption{Harmonized historical survey labor indicators, Bolivia, 1989-2005}
\label{tab:surveys}
\begin{adjustbox}{max width=\linewidth}
\begin{tabular}{lrrrrr}
\toprule
Period & Persons & Employment 14+ & Unemployment 14+ & Informality proxy & Labor-income Gini \\
\midrule
1986-1990 & 43,336 & 53.0 & 5.0 & 34.3 & 0.945 \\
1991-1995 & 25,933 & 53.8 & 2.7 & 36.2 & 0.910 \\
1996-2000 & 39,481 & 62.0 & 0.7 & 27.3 & 0.851 \\
2001-2005 & 26,374 & 60.1 & 1.9 & 35.0 & 0.692 \\
\bottomrule
\end{tabular}
\end{adjustbox}
\vspace{1mm}\begin{minipage}{0.95\linewidth}\footnotesize Survey indicators are mechanisms, not official poverty estimates. Persons may reflect effective harmonized person counts from the survey-processing pipeline.\end{minipage}
\end{table}

%% file: tables/tbl_diagnostics_clean.tex
\begin{table}[!htbp]
\centering
\scriptsize
\caption{Re-estimated Bolivia heterogeneous IV diagnostics}
\label{tab:diagnostics}
\begin{adjustbox}{max width=\linewidth}
\begin{tabular}{lrrrrrrr}
\toprule
Outcome & N & Countries & Beta +10 & HC3 SE & HC3 p & DW & BP p \\
\midrule
Poverty USD 4.20/day & 357 & 22 & 4.42 & 1.08 & <0.001 & 0.84 & <0.001 \\
Poverty USD 3/day & 357 & 22 & 3.57 & 1.13 & 0.002 & 1.08 & <0.001 \\
Poverty USD 8.30/day & 357 & 22 & 7.37 & 0.90 & <0.001 & 0.72 & <0.001 \\
Gini & 357 & 22 & 3.92 & 0.80 & <0.001 & 1.03 & <0.001 \\
\bottomrule
\end{tabular}
\end{adjustbox}
\vspace{1mm}\begin{minipage}{0.95\linewidth}\footnotesize Durbin-Watson below 2 indicates positive serial correlation. The Breusch-Pagan p-values indicate heteroskedasticity, so robust or clustered inference is required.\end{minipage}
\end{table}

%% file: tables/tbl_vif_clean.tex
\begin{table}[!htbp]
\centering
\small
\caption{Variance inflation factors for baseline controls before fixed effects}
\label{tab:vif}
\begin{adjustbox}{max width=\linewidth}
\begin{tabular}{lr}
\toprule
Control & VIF \\
\midrule
heritage overall & 1.06 \\
crec pib pc & 1.01 \\
inflacion cpi & 1.04 \\
terminos intercambio & 1.03 \\
\bottomrule
\end{tabular}
\end{adjustbox}
\end{table}

%% file: tables/tbl_adf_clean.tex
\begin{table}[!htbp]
\centering
\scriptsize
\caption{Stationarity diagnostics for Bolivia series}
\label{tab:adf}
\begin{adjustbox}{max width=\linewidth}
\begin{tabular}{lrrrr}
\toprule
Variable & N & ADF stat. & p & Stationary 5\% \\
\midrule
heritage overall & 30 & -3.46 & 0.009 & yes \\
efw & 30 & -2.23 & 0.197 & no \\
D efw & 29 & -3.92 & 0.002 & yes \\
poverty 420usd & 27 & -1.51 & 0.526 & no \\
D poverty 420usd & 26 & -1.95 & 0.310 & no \\
poverty 3usd & 27 & -1.09 & 0.717 & no \\
poverty 830usd & 27 & -1.31 & 0.624 & no \\
gini & 27 & -0.95 & 0.769 & no \\
\bottomrule
\end{tabular}
\end{adjustbox}
\end{table}